\documentclass{aastex631}
\usepackage{hyperref} 
\usepackage{natbib}
\usepackage{textcomp}
\usepackage{gensymb}
\usepackage{xcolor}
\usepackage{amsmath}
\usepackage{comment}
\usepackage{arydshln}
\usepackage{graphics}
\usepackage{booktabs}

\begin{document}
\title{Unveiling a young thick disk in the Milky Way}
\author[0000-0001-5258-1466]{Jianhui Lian}
\affiliation{South-Western Institute for Astronomy Research, Yunnan University, Kunming, Yunnan 650091, People’s Republic of China}
\email{jianhui.lian@ynu.edu.cn (JL)}  

\author{Min Du}
\affiliation{School of Physics and Astronomy,China West Normal University, Nanchong, Sichuan 637009, People’s Republic of China}
\email{dumin@xmu.edu.cn (MD)}

\author{Shuai Lu}
\affiliation{School of Physics and Astronomy,China West Normal University, Nanchong, Sichuan 637009, People’s Republic of China}

\author{Bingqiu Chen}
\affiliation{South-Western Institute for Astronomy Research, Yunnan University, Kunming, Yunnan 650091, People’s Republic of China}
\author{Gail Zasowski}
\affiliation{Department of Physics \& Astronomy, University of Utah, Salt Lake City, UT 84112, USA}
\author{Zhaoyu Li}
\affiliation{Department of Astronomy, School of Physics and Astronomy, Shanghai Jiao Tong University, 800 Dongchuan Road, Shanghai 200240, PR China}
\author{Xiaojie Liao}
\affiliation{Department of Astronomy, School of Physics and Astronomy, Shanghai Jiao Tong University, 800 Dongchuan Road, Shanghai 200240, PR China}
\author{Chao Liu}
\affiliation{National Astronomical Observatories, Chinese Academy of Sciences, Beijing 100012, PR China}
\affiliation{University of Chinese Academy of Sciences, Beijing 100049, PR China}


\begin{abstract}
The thickness of a galaxy's disk provides a valuable probe of its formation and evolution history. Observations of the Milky Way and local galaxies have revealed an ubiquitous disk structure with two distinctive components: an old thick disk and a relatively young thin disk. The formation of this dual-disk structure and the mechanisms that develop the thickness of the disk are still unclear. Whether the disk thickness inherit from the birth environment or is established through secular dynamical heating after formation is under debate. 
In this work we identify a relatively young ($\sim$6.6 billion years old) geometric thick disk in the Milky Way, with a scale height of {$0.64$~kpc at {the Solar Circle}. This young thick component exhibits comparable thickness and flaring strength to the canonical old thick disk but is more radially extended and systematically younger.} We also identify thin disk components that formed before and after this young thick disk. Detailed analysis of the solar vicinity structure suggests that the young thick disk marks the onset of a new phase of upside-down disk formation. These findings strongly discount the role of secular dynamical heating and support a turbulent, bursty birth environment as the primary mechanism behind thick disk formation. The existence of two thick disk components suggests that the Milky Way has undergone at least two episodes of turbulent and bursty star formation, likely triggered by galaxy mergers.   
\end{abstract}

\section{Introduction}
A dual disk structure, characterized by two components with distinct thicknesses, has been found to be ubiquitous in local disk galaxies \citep{yoachim2006,comeron2018}, including the Milky Way \citep{gilmore1983}. Extensive studies have revealed that these two disk components differ not only in thickness but also in a variety of other properties. The thick component primarily consists of older stars with lower metallicity, enhanced $\alpha$ abundance, and hotter kinematics \citep{fuhrmann1998,reddy2006,haywood2013,pinna2019}. 
Despite the long-standing recognition of both thick and thin disks, the mechanisms underlying the formation of these distinct components remain debated. 

Two main hypotheses have been proposed. The traditional view suggests that disk thickness develops through secular dynamic heating, potentially driven by interactions with substructures within the disk or with neighboring galaxies \citep{villumsen1985,quinn1993,villalobos2008,yi2023}. In this case, the old thick disk evolves from an initially thin disk, gaining thickness over time through dynamical processes. This is supported by the monotonic age-velocity dispersion relationship with higher velocity dispersion in older populations observed in the Milky Way \citep{wielen1977,nordstrom2004}. Conversely, the alternative hypothesis argues that the thickness is a result of the initial formation conditions, positing that the old thick disk {in the high-redshift Universe forms already thick within a turbulent environment \citep{brook2004,bournaud2009} or thickens rapidly via short-timescale dynamical heating in a disc with high gas fraction \citep{bland2024,bland2025}.} This scenario is favored by the observations of high-redshift galaxies, which generally exhibit turbulent kinematics \citep{elmegreen2005,weiner2006,ubler2019} and geometrically thick disks \citep{elmegreen2006,hamilton2023,lian2024a}. 

The Milky Way, as currently the only galaxy that allows temporally-resolved observations of stellar populations, offers unique insights into the formation and evolution of the galaxy disk thickness. While large-area photometric surveys have characterized the overall stellar structure of the Milky Way \citep{juric2008,bland2016}, its temporal evolution is more effectively illuminated by large-scale stellar spectroscopic surveys, such as APOGEE \citep{majewski2017},
GALAH \citep{martell2017}, and LAMOST \citep{zhao2012}. These surveys detail mappings of a large number of individual stars born at different epochs on a galaxy scale. Leveraging the wide spatial coverage of these surveys, many works have explored the structure of mono-abundance/age populations, which reflects the growth history of the Milky Way \citep{bovy2012b,bovy2016,mackereth2017,yu2021,lian2022b}.

In this work, based on data from the APOGEE spectroscopic survey \citep{majewski2017} in combination with Gaia astrometric data \citep{gaia2016}, we investigate the vertical structure of mono-abundance populations of the Milky Way and identify an intriguingly relatively young disk population in the Milky Way that exhibits a thickness comparable to the canonical old thick disk {at a given radius}. 
This finding imposes strong constraints on the formation mechanisms of the thick disk. 

\section{Data}

\begin{figure*}
	\centering
	\includegraphics[width=20cm,viewport=150 30 2500 1400,clip]{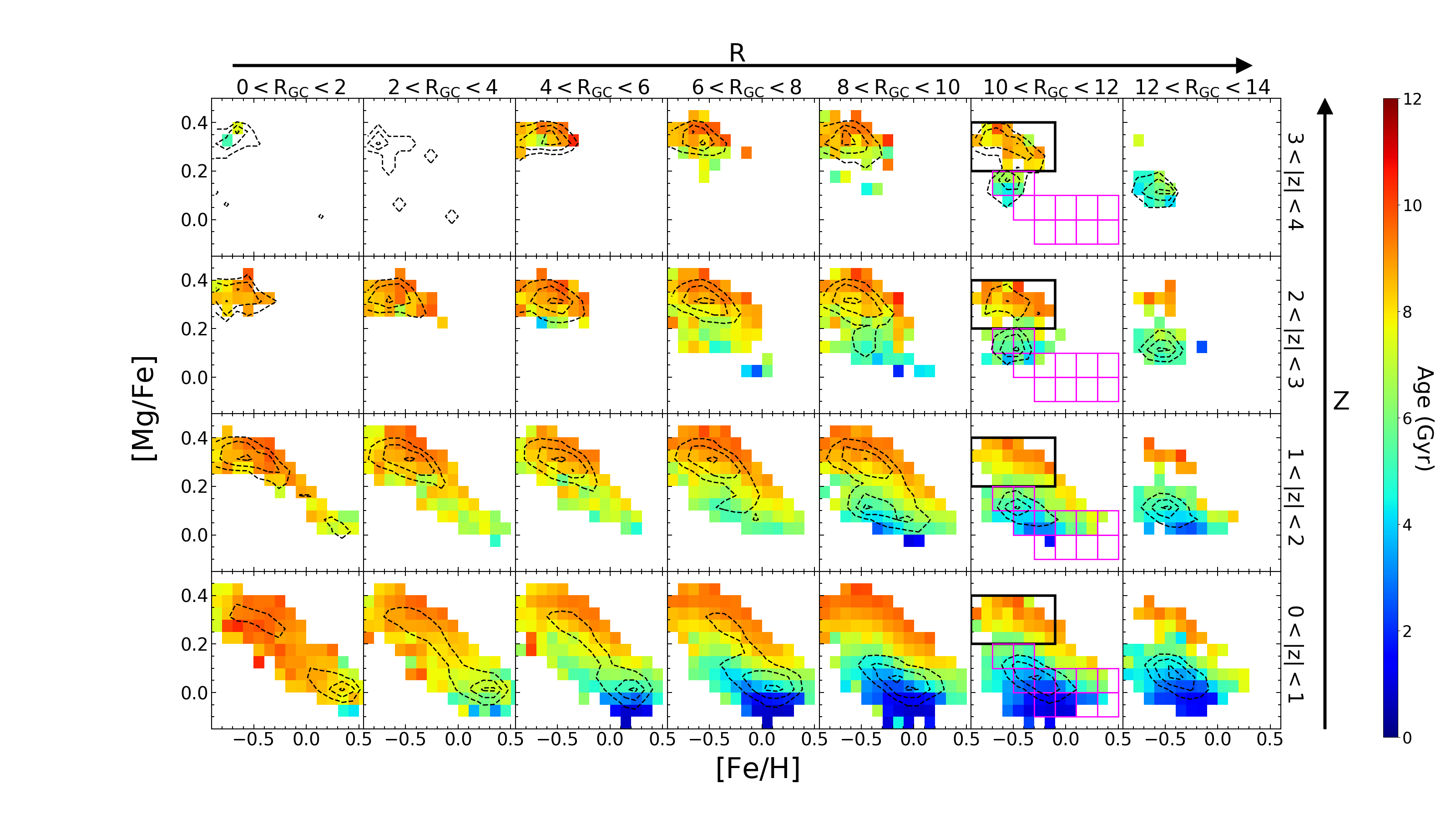}
	\caption{Averaged age map of APOGEE stars in the [Mg/Fe]-[Fe/H] diagram across the Galaxy. [Mg/Fe]-[Fe/H] distributions are examined within the ranges of ${\rm -0.9<[Fe/H]}<0.6$ and ${\rm -0.15<[Mg/Fe]<0.5}$, as a function of radius (${\rm 0<R<14\ kpc}$ in horizontal direction) and height (${\rm |Z|<4\ kpc}$ in vertical direction). Dashed contours outline the density distribution at level of 20\%, 50\%, and 90\% of the peak density in each panel. Pink {and black} boxes in the sixth column highlight the {low- and high-$\alpha$} mono-abundance populations for which the scale height is presented in Figure~\ref{hz}.} 
	\label{mgfe-feh}
\end{figure*} 

The data usage and description in this work is following \citep{lian2024b}. We use the data from the final data release of APOGEE survey within the SDSS-IV project \citep{blanton2017,sdss-dr17}. APOGEE is a massive near-infrared, high-resolution spectroscopic survey \citep{majewski2017} that provides high-quality spectra with a signal-to-noise (S/N) ratio normally exceeding 50. It has delivered robust and precise stellar parameters and elemental abundances for $\sim$0.6 million stars throughout the Galactic disc, bulge, and halo \citep{zasowski2013,zasowski2017,beaton2021,santana2021}. 

The chemical abundances (e.g., [Fe/H], [Mg/Fe]) and stellar parameters (e.g., log(g) and T$_{\rm eff}$) for individual stars are measured from the spectra by custom pipelines that incorporate a new custom line list (ASPCAP) \citep{nidever2015,garcia2016,smith2021}. Additionally, stellar ages, spectro-photometric distances, and orbit parameters are adopted from the astroNN APOGEE value-added catalog, which employs a deep-learning approach to analyze spectroscopic data from APOGEE and asterometric data from Gaia \citep{mackereth2019,leung2019}. The typical age and distance uncertainties are 30\% and 10\%, respectively. 

To ensure reliable measurement of stellar parameters and chemical abundances, we select our sample of disk stars from the APOGEE main survey using the following set of criteria: 
\begin{itemize}
    \item spectra S/N$>50$;
    \item surface gravity log($g$)$<3.5$;
    \item orbit eccentricity $e<0.7$;
    \item target flag set to be the main survey (EXTRATARG==0 ).
\end{itemize}

Figure~\ref{mgfe-feh} displays the distribution of APOGEE stars in the [Mg/Fe]-[Fe/H] abundance diagram, color-coded by average age, as a function of radius and height. The [Mg/Fe]-[Fe/H] distribution shows a clear bimodality with two sequences of distinctive $\alpha$ element (e.g., Mg) abundances, known as $\alpha$-bimodality \citep{fuhrmann1998,reddy2006,haywood2013}. The relative densities of these two sequences vary substantially across the Galaxy \citep{hayden2015,queiroz2018}, suggesting a strong dependence of the disk structure on chemical abundances. Specifically, the high-$\alpha$ sequence exhibits a larger scale height and extends further in the vertical direction compared to the low-$\alpha$ sequence \citep{bovy2012b,hayden2015}. This distinction leads to the classification of the high- and low-$\alpha$ sequences as chemically defined thick and thin disks, respectively. 

\section{Results and Discussion}
\subsection{Identification of a young thick disk component}

\begin{figure*}
	\centering
	\includegraphics[width=\textwidth]{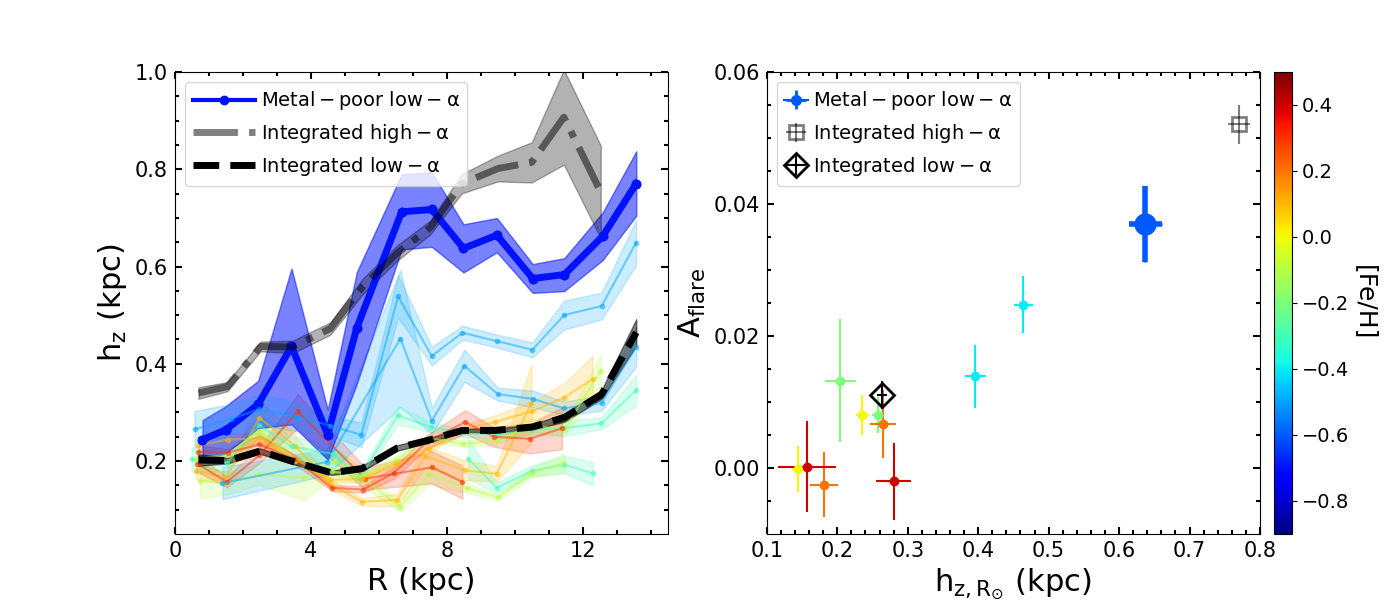}
	\caption{Vertical structure of mono-abundance and integrated populations of the Milky Way. {\sl Left:} scale height as a function of radius for various disk components. The gray dash-dotted and black dashed lines indicate the scale height radial profiles of integrated high- and low-$\alpha$ populations, {marked by the black and pink boxes in Figure~\ref{mgfe-feh}}, respectively. The colourful lines represent the radial profiles of low-$\alpha$ populations, split into mono-abundance space in [Mg/Fe] and [Fe/H], color-coded by [Fe/H]. {Note that the vertical axis spans only $\sim$7\% of the distance range of the horizontal axis.} {\sl Right:} Flaring strength (${\rm A_{flare}}$) versus scale height at solar {Galactocentric} radius for high- and low-$\alpha$ populations. The flaring strength is represented by the best-fitted slope of the scale height radial profile. The metal-poor low-$\alpha$ population emphasized in the main paper is marked with a thickened line in the left-hand panel and an enlarged circle in the right. Shade regions and error bars represents 1$\sigma$ uncertainties of the estimates of flaring strength and scale height.}
	\label{hz}
\end{figure*} 

\begin{figure*}
	\centering
	\includegraphics[width=\textwidth]{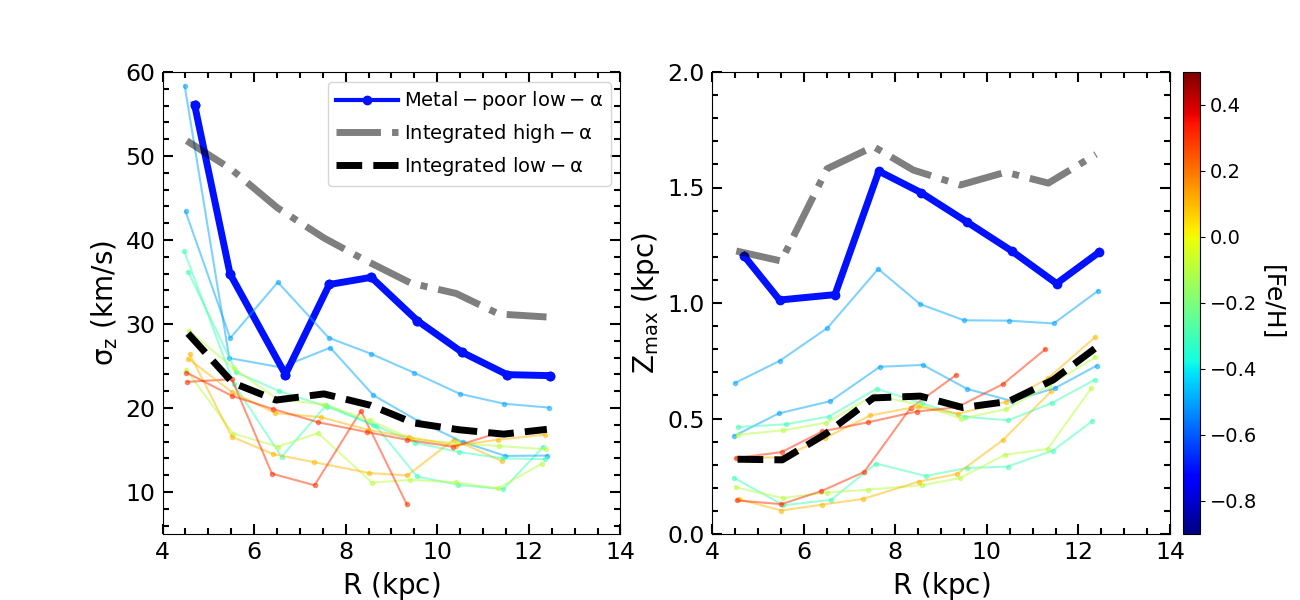}
	\caption{Vertical velocity dispersion and maximum orbit height of integrated high- and low-$\alpha$ and mono-abundance low-$\alpha$ populations. Raw sample from the APOGEE survey is used here without correcting for the selection function. {Note that the vertical axis spans only 20\% of the distance range of the horizontal axis.}             } 
	\label{orbit}
\end{figure*} 

One feature that has not been fully appreciated in previous studies is the existence of two groups of stars with distinctive age and $\alpha$ abundance at a large distance off the disk plane (${\rm |z|>3}$~kpc) and beyond the solar {Galactocentric} radius (${\rm 8<R<14}$~kpc). The first group consists of the expected canonical old thick disk stars, characterized by low metallicity and enhanced $\alpha$ abundance (${\rm [Fe/H]\sim-0.6\ dex}$, [Mg/Fe]$\sim$0.3 dex). The second group, while similarly low in metallicity, exhibits {a} systematically younger age and lower $\alpha$ abundance ([Mg/Fe]$\sim$0.15 dex). Both groups exhibit disk-like orbits but with relatively hot kinematics, characterized by high vertical velocity dispersion and maximum orbit height. 

While sharing consistent vertical structure and kinematics, they are distinctly separated in age and $\alpha$ abundance, with a clear density dip at [Mg/Fe] $= 0.2$ dex. The median ages of the high-$\alpha$ and metal-poor (MP) low-$\alpha$ populations are 9.3~Gyr and 6.6~Gyr, respectively. In addition, the fraction of the MP low-$\alpha$ population at large height increases rapidly towards larger radii and {becomes} the predominated thick component beyond ${\rm R>12}$~kpc, {leading to a clear negative age gradient at large height above the disc plane \citep{martig2016}.} In light of these observational results, we hypothesize that the relatively young, MP low-$\alpha$ population may represent a new, young thick disk substructure, comparable in thickness to the canonical old thick disk represented by the high-$\alpha$ population. 

{To further investigate this possibility, we divide the disk into mono-abundance populations (MAPs) defined in terms of [Fe/H] and [Mg/Fe], reconstructe their intrinsic density distribution, and measure the scale height, which represents the disk thickness, as a function of radius. These mono-abundance populations are marked in Figure~\ref{mgfe-feh}, with {a} uniform interval of 0.2~dex in [Fe/H] and 0.1~dex in [Mg/Fe]. 
We treat the high-$\alpha$ disk as a cohesive entity and select the high-$\alpha$ and metal-poor low-$\alpha$ stars with the following selection criteria.
\begin{itemize}
    \item High-$\alpha$: ${\rm -0.9<[Fe/H]<-0.1\ dex}$ and ${\rm 0.2<[Mg/Fe]<0.4\ dex}$;
   \item Metal-poor (MP) low-$\alpha$: ${\rm -0.7<[Fe/H]<-0.5\ dex}$ and ${\rm 0.1<[Mg/Fe]<0.2\ dex}$;
\end{itemize}} 

To measure the {intrinsic density distribution and} scale height of these MAPs, we follow the procedure in \citet{lian2022b}. { For each mono-abundance population, We first use up--to--date PARSEC stellar isochrones \citep{bressan2012} and a 3D dust extinction map \citep{bovy2015a}, sampling with a Kroupa initial mass function \citep{kroupa2001}, to generate mock stellar catalogs and simulate stellar distribution in the color-magnitude diagram at different distances of each field targeted by the APOGEE survey.} We then apply the APOGEE selection criteria, which is based on the 2MASS $J-K_{\rm s}$ and $H$ color-magnitude diagram \citep{zasowski2013,zasowski2017,beaton2021,santana2021}, to estimate the probability of a star being observed by APOGEE, i.e., the selection function of the survey. {PARSEC isochrones do not have various $\alpha$ abundance options other than the solar value. However, [$\alpha$/Fe] has a much weaker effect on the location of stars in the color-magnitude diagram than their distance from Earth, line-of-sight extinction, age, and [Fe/H], we expect that using isochrones without various $\alpha$ abundance options does not significantly affect the estimate of the survey selection function and the results of this work. }

By dividing the observed number density of stars in the APOGEE by {the probability of being observed}, we derive the intrinsic 3D stellar density distribution. Finally, at each radius, we fit a sech$^2$ function 
\begin{equation}
    {\rho=\rho_0\times\frac{4}{{(e^{\Delta z/2h_{z}}+e^{-\Delta z/2h_{z}})}^2}}
\end{equation}
to the density distribution perpendicular to the disk plane using curve\_fit function in Python Scipy package to obtain the scale height of each MAP. Here $h_{\rm z}$ represents the scale height. The uncertainties of the scale height are estimated using a Monto Carlo simulation approach. We first generate a new set of randomized density distributions based on the corresponding density error. For each new set of density distribution, we derive a new scale height measurement. By repeating this process 100 times, we compute the standard deviation of the 100 measurements, which serves as the uncertainty of our scale height measurements. These processes are conducted independently for each MAPs. 

The obtained scale heights as a function of the radius of these MAPs are illustrated in Figure~\ref{hz}. {Consistent with the raw density distribution in Figure~\ref{mgfe-feh}, the MP low-$\alpha$ population exhibits vertical structure similar to the high-$\alpha$ population and distinct from the rest of low-$\alpha$ populations.}
Within the inner Galaxy (${\rm R<5}$~kpc), the scale heights of all MAPs are overall small, with the MP low-$\alpha$ population positioned between the high-$\alpha$ and {the rest of} low-$\alpha$ populations. Beyond the inner Galaxy, the scale heights of the high-$\alpha$ and MP low-$\alpha$ populations both increase rapidly, reaching 0.77 and 0.64~kpc around solar {Galactocentric} radius at 8.2~kpc, respectively. In contrast, the {majority of} low-$\alpha$ populations exhibit a nearly constant scale height of $\sim$0.2~kpc, which is typical for the Galactic thin disk \citep{bland2016}.  
Outside the {Solar Circle}, the scale height of MP low-$\alpha$ stars remains high and largely invariant, while that of high-$\alpha$ stars continues to increase.
{The increase of disk thickness with radius is commonly referred to as disk flaring \citep{lopez-corredoira2002,momany2006,minchev2015}. For simplicity, we fit the scale height radial profile with a linear model and use the slope to indicate the strength of flaring (${\rm A_{flare}}$).}
Overall, the high-$\alpha$ and MP low-$\alpha$ populations constitute a resembling substructure that is substantially thicker at a given radius and at the same time exhibits stronger flaring compared to the thin disk represented by the {majority of} low-$\alpha$ populations. This finding supports the existence of a young thick disk in the Milky Way, in addition to the well-established old thick disk. The coexistence of the two MAPs with thick disk-like structure, along with their distinct ages {and $\alpha$ abundance}, implies that our Galaxy has undergone at least two episodes of thick disk formation. 

The thickness of a disk in equilibrium is expected to be positively correlated with the velocity dispersion in the vertical direction and the maximum height of the stellar orbit \citep{Sotillo-Ramos2023}. Figure~\ref{orbit} shows the vertical velocity dispersion and maximum orbit height of the MAPs investigated in this work. Consistent with the scale height measurement, the MP low-$\alpha$ population shows large vertical velocity dispersion and maximum orbit height that are comparable to the high-$\alpha$ populations and systematically higher than other low-$\alpha$ populations. This serves as independent evidence supporting the thick nature of the MP low-$\alpha$ population.  

{While the MP low-$\alpha$ population comprises a geometrically thick disk structure, it is embedded in the low-$\alpha$ disk, which is overall thin and flares weakly within $R<12$~kpc. The geometric thick nature disappear when analyzing the integrated or mono-age populations of the low-$\alpha$ disk, thus being overlooked in previous studies of the vertical structure of the Galactic disk\citep{mackereth2017,ting2019}. Abundance information is essential to correctly identify this intriguing substructure.} {Although signatures of the young thick disk have been presented in previous works \citep{hayden2015,martig2016}, we for the first time derive its structure parameters that allow us to directly compare this population with the old canonical thick disk and the rest of the low-$\alpha$ disk to confirm its geometric thick nature and to appreciate its outstanding position among low-$\alpha$ populations and similarity to high-$\alpha$ population in vertical structure.}

{The Galactic disk has a decreasing surface density, and thus shallower potential well in the vertical direction, at larger radii. As a result, stars at larger radii can move further off the disk plane, presenting a flaring feature. This may explain the relatively weak flaring observed in the majority of low-$\alpha$ populations. However, the much higher flaring strength of the young thick disk population is likely not solely due to the declining disk potential, implying the contribution from a different process (e.g., galaxy interactions) or the mechanism acting under a different condition. A detailed calculation of the flaring caused by the declining potential would be interesting but is beyond the scope of this paper.}

\subsection{Broken age-scale height and age-velocity dispersion relationships}
\begin{figure*}
	\centering
	\includegraphics[width=21cm,viewport=130 10 2000 530,clip]{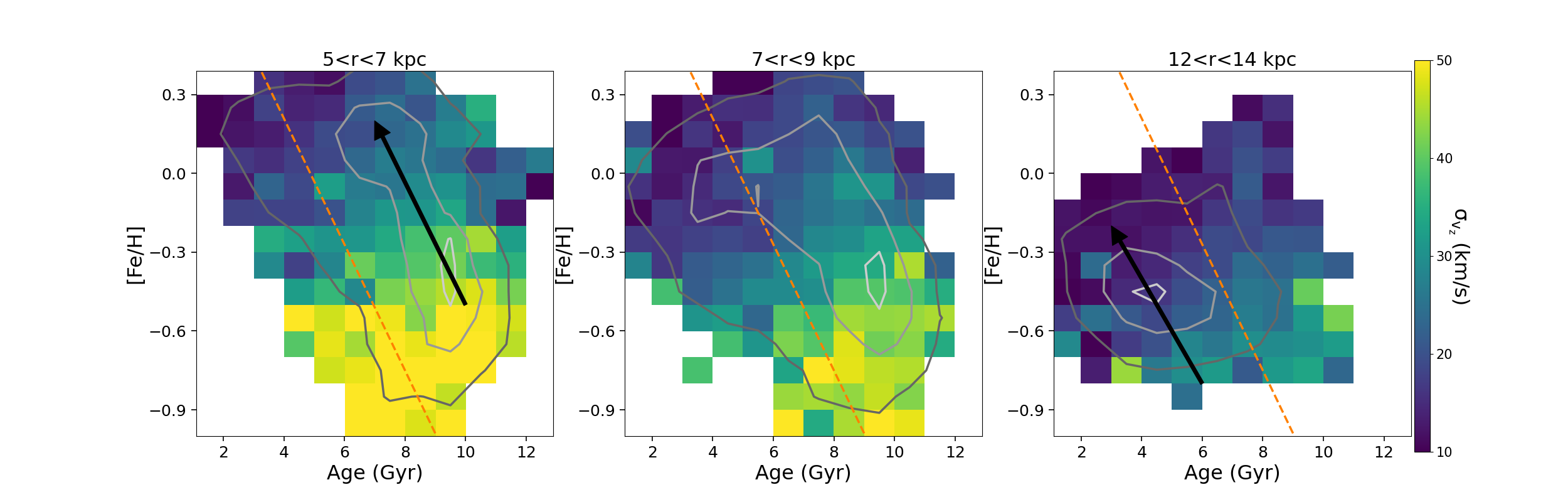}
	\caption{Distribution of vertical velocity dispersion in age-[Fe/H] plane at three radii. Raw APOGEE sample is used to calculate the velocity dispersion. Grey contours in each panel outline the density distribution with level of 30\%, 60\%, and 90\% of the peak density. Black arrows in the left and right panels indicate the trend of decreasing vertical velocity dispersion along the two age-metallicity sequences that are predominated in the inner and outer Galaxy, respectively. Orange dashed line is used to separate the two age-metallicity sequences.                
	} 
	\label{vdisp-age-feh}	
\end{figure*} 

\begin{figure*}
	\centering
	\includegraphics[width=\textwidth]{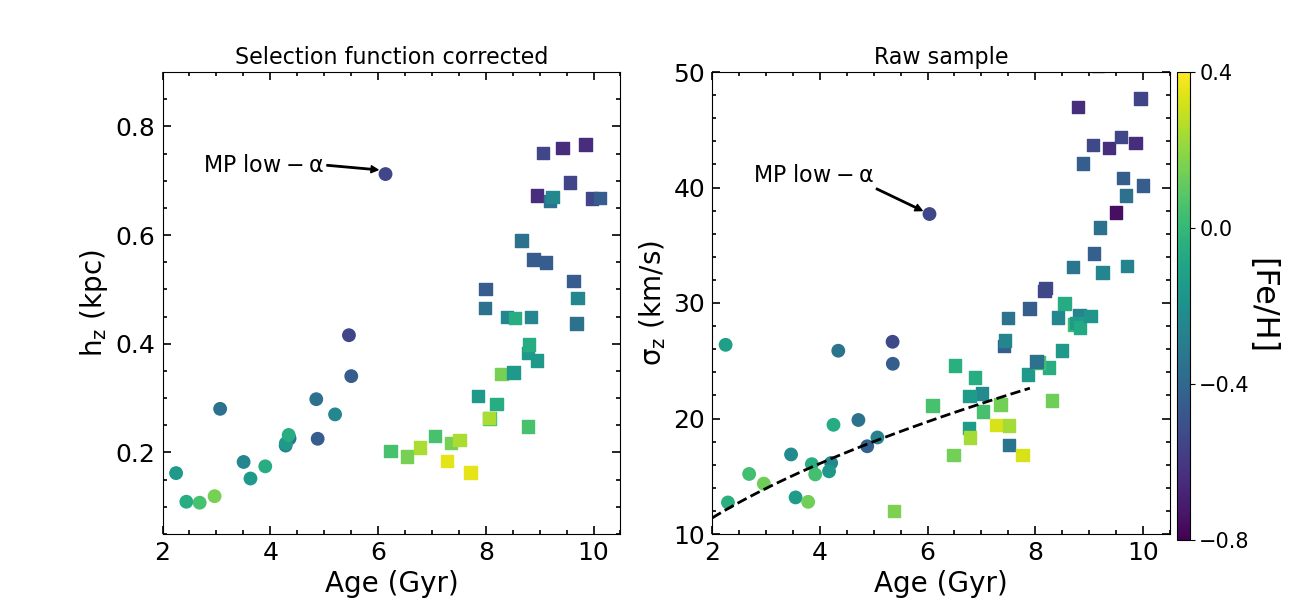}
	\caption{Scale height and vertical velocity dispersion as a function of stellar age for mono-abundance populations (MAPs). MAPs along the old and young age-metallicity sequences are indicated as squares and circles, respectively. The separation of these two sequences is defined by a straight line in the age-metallicity plane. Color coding reflects the metallicity of each MAP. The scale height is calculated from the intrinsic density distribution of MAPs after correcting for the selection function, whereas the velocity dispersion is derived from the raw sample data. Dashed curve in the right panel represents the power law with a slope of 0.5, illustrating the predictions of secular heating scenario \citep{martig2014}.} 
	\label{hzv}
\end{figure*} 

An interrupted age--metallicity relation in the solar vicinity has been suggested in recent massive stellar spectroscopic surveys \citep{anders2017,feuillet2018,silva2018,hasselquist2019,lian2020a,xiang2022,imig2023}. More recent observations with wider radial coverage and more precise age determinations reveal that the interrupted age--metallicity relation is the result of two coexisting age--metallicity sequences with systematic offset in age or metallicity \citep{lian2022a,xiang2022}. The old sequence is predominated in the inner Galaxy with a higher metallicity at a given age than the younger sequence that is predominated in the outer Galaxy. 

Figure~\ref{vdisp-age-feh} shows the age--[Fe/H] distribution of APOGEE stars in three radial bins color coded by vertical velocity dispersion ($\sigma_{\rm z}$). The two age--metallicity sequences are predominated in the inner and outer radial bins, respectively, while coexisting {at the Solar Circle}. To separate these two sequences we apply a straight line in age--[Fe/H] plane in light of the age--[Fe/H] distributions in the inner and outer Galaxy as following:
\begin{equation}
    {\rm [Fe/H]} = -0.24*Age+1.17. 
\end{equation}
The trend of decreasing $\sigma_{\rm z}$ with time along each age--metallicity sequence is indicated by the black arrows in the left and right panels. 

The observed number density of APOGEE stars is substantially higher {in the solar neighborhood} compared to that at greater distance, allowing a detailed investigation of the vertical structure. Consequently, we select a subsample of stars within the range of $7<R<9$~kpc and define mono-abundance populations with a finer grid of 0.1~dex in [Fe/H] and 0.05~dex in [Mg/Fe]. This grid is half the resolution of the MAPs defined to study the global structure presented in Figure~2. The same approach as in \textsection3.1 has been adopted to derive the scale height of these MAPs.  
Figure~\ref{hzv} illustrates the scale height and vertical velocity dispersion ($\sigma_{\rm z}$) as a function of the mean age of the MAPs at the solar {Galactocentric} radius. 

{A monotonic age-velocity dispersion relation has been established for decades and interpreted as a consequence of secular vertical heating \citep{wielen1977,nordstrom2004} . In this work, by carefully segregating disk populations based on chemical abundances, we uncover a dramatic discontinuity in the age-velocity dispersion relation that disrupts the monotonic trend. This finding is further supported by a similar discontinuity observed in the scale height. The MP low-$\alpha$ population exhibits a substantially higher scale height and velocity dispersion compared to the metal-rich low-$\alpha$ disk population that formed immediately before.}

Inspired by the existence of two disconnected age-[Fe/H] evolutionary sequences in the solar {neighborhood}~\citep{lian2022a,xiang2022}, we mark the MAPs following {the old and young} age-[Fe/H] sequences according to their average age based on Eq(2) with {squares and circles in Figure~\ref{hzv}, respectively}. Remarkably, these disconnected two age-[Fe/H] sequences also manifest in the age-scale height and age-velocity dispersion planes, revealing two coherent sequences of chemical enrichment and structural evolution: younger, more metal-rich populations are found within a thinner and colder disk structure. 
{The MP low-$\alpha$ population signifies the onset of the second evolutionary sequence.}
{While the gradual increase in scale height and velocity dispersion observed in older populations can be attributed to secular vertical heating \citep{martig2014,ting2019}, the pronounced excess found in the MP low-$\alpha$ and high-$\alpha$ populations contradicts the disk heating theory, suggesting that they originated in a turbulent environment. Furthermore, the rapid decrease in scale height and velocity dispersion of populations formed immediately after these two thick disk populations indicates two distinct phases of upside-down disk formation.}

The formation of metal-rich low-$\alpha$ thin disk populations situated between the two thick disk populations provides another compelling evidence against {dynamical heating} as the primary mechanism for thick disk formation. If this were the case, the preexisting thin disk would expect to be dynamically heated as well during the formation of the young thick disk. Instead, our findings suggest that birth conditions play a crucial role in determining disk thickness.  
The high-$\alpha$ abundance and low metallicity of the old thick disk indicate that it likely formed during an early starburst, possibly linked to the Gaia Encaledus merger \citep{robin2014,grand2018,buck2020,lian2022b}. Observations of high-redshift galaxies have revealed turbulent kinematics \citep{elmegreen2005,weiner2006,ubler2019} and the presence of geometrically thick disks \citep{elmegreen2006,hamilton2023,lian2024a}. Together, these pieces of evidence suggest that the old thick disk of the Milky Way was born already thick from a turbulent and bursty environment. 

It is reasonable to suspect that the formation of the young thick disk is connected to a starburst possibly triggered by a recent galaxy merger event. Recently, there has been growing evidence suggesting that the Milky Way may have experienced a recent starburst approximately 5-6 billion years ago \citep{ruiz-lara2020}, likely associated with the first passage of the Sagittarius dwarf galaxy \citep{laporte2018}. Detailed chemical evolution modelling further supports the hypothesis that the MP low-$\alpha$ population formed during this starburst, in conjunction with a metal-poor gas accretion and dilution event likely supplied by the Sagittarius dwarf galaxy \citep{spitoni2019,lian2020a}. Compared to the old thick disk, the young thick disk is {slightly less thick} but more radially extended. This indicates that the recent starburst responsible for the formation of the young thick disk was relatively weaker and primarily occurred at larger radii compared to the early burst that formed the old thick disk {in the inner Galaxy}.  


\section{Conclusion}
In this work, based on the data from large-scale stellar spectroscopic and astrometric surveys, we identified a relatively young ($\sim$6.6~Gyr old) thick disk in the Milky Way that formed several billion years after the canonical old thick disk and following a well-defined thin disk. This young thick disk shares comparable scale height {at a given radius and flaring strength} with the old thick disk (scale height of 0.64 {vs.} 0.77~kpc at solar {Galactocentric} radius) but exhibits a lower $\alpha$-enhancement and a more extended radial distribution. Its presence causes a discrepancy between the geometric and chemical definitions of thick disk {in the literature}. Our detailed analysis at the solar {Galactocentric} radius further reveals non-monotonic evolution of scale height and velocity dispersion, which contradicts traditional heating scenario but instead supports a scenario in which the thick disk is born thick from a turbulent and bursty environment. 

Our findings reveal a multiphase evolution history of the Milky Way's disk that encompasses two episodes of `upside-down' disk formation. Specifically, the two thick disk populations mark the onset of these two phases of upside-down disk formation. This study presents critical observational results that pin down the primary mechanisms governing thick disk formation and reveal an intriguing connection between thick disk formation and the bursty star formation histories of galaxies, particularly in relation to galaxy merger and gas accretion events. These insights are poised to largely reshape our understanding of the disk formation and evolution history of the Milky Way, as well as galaxies in general. 

Data presented in this work are available in the public repository \url{https://github.com/lianjianhui/source-date-for-young-thick-disk-paper.git}.

\section*{Acknowledgements}
We thank Misha Haywood, Nicholas Martin, Zheng Yuan, Maosheng Xiang, and Yang Huang for insightful discussions. 
J.L. acknowledges support by National Natural Science Foundation of China (No. 12473021), National Key R\&D Program of China (No. 2024YFA1611600), Yunnan Province Science and Technology
Department Grant (No. 202105AE160021 and 202005AB160002), and the Start-up Fund of Yunnan University (No. CY22623101). M.D. acknowledges the support by the
Fundamental Research Funds for the Central Universities (No. 20720230015) and the Natural Science Foundation of Xiamen, China (No. 3502Z202372006).

Funding for the Sloan Digital Sky Survey IV has been provided by the Alfred P. Sloan Foundation, the U.S. Department of Energy Office of Science, and the Participating Institutions. SDSS-IV acknowledges
support and resources from the Center for High-Performance Computing at the University of Utah. The SDSS web site is www.sdss.org.

SDSS-IV is managed by the Astrophysical Research Consortium for the 
Participating Institutions of the SDSS Collaboration including the 
Brazilian Participation Group, the Carnegie Institution for Science, 
Carnegie Mellon University, the Chilean Participation Group, the French Participation Group, Harvard-Smithsonian Center for Astrophysics, 
Instituto de Astrof\'isica de Canarias, The Johns Hopkins University, Kavli Institute for the Physics and Mathematics of the Universe (IPMU) / 
University of Tokyo, the Korean Participation Group, Lawrence Berkeley National Laboratory, 
Leibniz Institut f\"ur Astrophysik Potsdam (AIP),  
Max-Planck-Institut f\"ur Astronomie (MPIA Heidelberg), 
Max-Planck-Institut f\"ur Astrophysik (MPA Garching), 
Max-Planck-Institut f\"ur Extraterrestrische Physik (MPE), 
National Astronomical Observatories of China, New Mexico State University, 
New York University, University of Notre Dame, 
Observat\'ario Nacional / MCTI, The Ohio State University, 
Pennsylvania State University, Shanghai Astronomical Observatory, 
United Kingdom Participation Group,
Universidad Nacional Aut\'onoma de M\'exico, University of Arizona, 
University of Colorado Boulder, University of Oxford, University of Portsmouth, 
University of Utah, University of Virginia, University of Washington, University of Wisconsin, 
Vanderbilt University, and Yale University.

\bibliographystyle{aasjournal}
\bibliography{main}{}

\end{document}